\documentclass[10pt,conference]{IEEEtran}
\usepackage{epsfig,setspace,amsmath,epsf,amssymb,bm,theorem,cite, graphicx, epstopdf, algorithm, algpseudocode,float,color,bbm}
\usepackage{multirow}
\usepackage{authblk}
\usepackage[table,xcdraw]{xcolor}
\usepackage{mathtools}

\newtheorem{theorem}{Theorem}

\newtheorem{remark}{Remark}
\newtheorem{lemma}{Lemma}

\newenvironment{Proof}[1]{\medskip\par\noindent{\bf Proof:\,}\,#1}{{\mbox{\,$\blacksquare$}\par}}

\IEEEoverridecommandlockouts

\allowdisplaybreaks

\begin{document}

\title{Asymmetric $X$-Secure $T$-Private Information Retrieval: More Databases is Not Always Better}

\author{Mohamed Nomeir \qquad Sajani Vithana \qquad Sennur Ulukus\\
	\normalsize Department of Electrical and Computer Engineering\\
	\normalsize University of Maryland, College Park, MD 20742 \\
	\normalsize \emph{mnomeir@umd.edu} \qquad \emph{spallego@umd.edu} \qquad \emph{ulukus@umd.edu}}

\maketitle

\begin{abstract}
   We consider a special case of $X$-secure $T$-private information retrieval (XSTPIR), where the security requirement is \emph{asymmetric} due to possible missing communication links between the $N$ databases considered in the system. We define the problem with a communication matrix that indicates all possible communications among the databases, and propose a database grouping mechanism that collects subsets of databases in an optimal manner, followed by a group-based PIR scheme to perform asymmetric XSTPIR with the goal of maximizing the communication rate (minimizing the download cost). We provide an upper bound on the general achievable rate of asymmetric XSTPIR, and show that the proposed scheme achieves this upper bound in some cases. The proposed approach outperforms classical XSTPIR under certain conditions, and the results of this work show that unlike in the symmetric case, some databases with certain properties can be dropped to achieve higher rates, concluding that more databases is not always better. 
\end{abstract}

\section{Introduction}

Private information retrieval (PIR) \cite{chor} describes a fundamental privacy setting where a user wishes to download a required message from a set of $K$ messages stored in a system of $N$ non-colluding databases without revealing the required message index to any of the databases. The capacity of PIR was characterized in \cite{c_pir}, after which a number of  variants of the basic PIR problem was investigated to obtain fundamental performance bounds and achievable schemes \cite{c_spir, wang_spir, tspir_mdscoded, banawan_pir_mdscoded, codedstorage_adversary_tpir, Salim_CodedPIR, Kumar_PIRarbCoded, banawan_multimessage_pir, multimessage_pir_sideinfo, colluding, byzantine_tpir,  uncoded_constrainedstorage_pir, batuhan_hetero, utah_hetero, wei_banawan_cache_pir, chaoTian, semantic_pir, salim_singleserver_pir}. 

PIR with secure storage is one such variant that considers the security of the stored messages against the databases in which they are stored, in addition to the privacy requirement of the message index present in basic PIR. In the model presented in \cite{first_xsecure} for PIR with secure storage, any collection of up to $X$ databases are allowed to communicate and still obtain no information about the stored messages. This is known as the $X$-security constraint on the stored messages. The model presented in \cite{csa} considers $X$-security as well as $T$-privacy, which allows up to any $T$ databases to collude and still not learn any information on the user-required message index. This is known as $X$-secure $T$-private information retrieval (XSTPIR). In \cite{csa}, the asymptotic capacity of XSTPIR for arbitrary $X$, $T$, and $N$ is characterized as,
\begin{align}
\label{asymptotic case}
\lim_{K \rightarrow \infty }C_{XSTPIR}(X\!,T\!,N\!,K) = \begin{cases}
1-\frac{X+T}{N},\!\!\! &\quad  X+T < N,\\
0, \!\!\!&\quad X+T \geq N.
\end{cases}
\end{align}

In practice, the databases that store these messages are owned by different data storage companies such as Amazon Web Services, Dell, IBM, etc. Based on their requirements and policies, it may not be that any $X$ of them are always allowed to communicate with each other. Consider a scenario where there is only one specific group of $X$ databases that are able to communicate to learn about the stored messages, and all other possible communications among the databases only support the data exchange of less than $X$ databases. Based on the existing schemes, the security constraint in this case is satisfied by considering $X$-security, to handle the worst case, i.e., the one group of $X$ databases that communicate. However, the component of $X$-security that ensures data-security against \emph{any} group of $X$ databases is not necessary in this case as no other group of $X$ databases are allowed to communicate. As the asymptotic capacity in \eqref{asymptotic case} decreases with increasing $X$, using existing schemes that are developed for \emph{symmetric} cases, i.e., all groups of $X$ communicating databases, is not optimal for cases with asymmetric database communications.  

In this paper, we consider asymmetric communications among databases, defined by a given set of arbitrary communication links specified for the given system of databases; as in \cite{nan_eaves}. As the symmetric $X$-security assumption deteriorates the overall rate in asymmetric settings, we propose a novel database grouping approach followed by a data encoding structure that ensures privacy against any $T$ colluding databases and security against any communications defined among the databases in the system. We call this \emph{asymmetric} XSTPIR (A-XSTPIR). As a converse result, we provide an upper bound on the general achievable rate of A-XSTPIR, and show that the proposed scheme achieves this upper bound in some cases.

In this work, we show that for a given set of communication links, the proposed A-XSTPIR scheme achieves higher rates compared to classical XSTPIR in certain settings. Furthermore, the proposed approach provides a systematic method to drop the highly communicative databases in some cases to improve the overall rate. This implies that having more databases is not always better in a setting with asymmetric database communications, which is counter-intuitive from the perspective of classical XSTPIR.  

\section{System Model and Problem Formulation} \label{sysmodel}

We consider a system of $N$ databases out of which any $T$ can collude to obtain information about the user-required message index. Each database contains the same set of $K$ independent and identically distributed messages generated from the field $\mathbb{F}_q$, denoted by $(W_1,\ldots,W_K)=W_{1:K}$, of equal length $L$. Thus,
\begin{align} \label{2}
    H(W_{1:K})= \sum_{i=1}^{K} H(W_i) = KL.
\end{align}
All possible communication links among the databases that are used to share information on the stored data to learn about the stored messages is given by the communication matrix $B_X$, of size $N \times M$, where $M$ is the number of links. For example, if the first communication link specifies the communications between databases $i_1$, $i_2$, and $i_3$, then $B_X(i_1,1)=B_X(i_2,1)=B_X(i_3,1)=1$, and the rest of the entries in the first column of $B_X$ are zero. The messages $W_{1:K}$ need to be secure against any communications between databases defined by $B_X$, i.e.,
\begin{align} \label{3}
    I(W_{1:K}; S_{\mathcal{X}_i})= 0, \quad i\in\{1,\dotsc,M\},
\end{align}
where $S_{\mathcal{X}_i}$ is the collection of stored data of all databases in the $i$th communication link, $\mathcal{X}_i$. A user who wants to retrieve $W_k$, where $k$ is chosen uniformly at random, sends a query to each database $n$ denoted by $Q_{n}^{[k]}$. All queries sent by the user to the system of databases is given by $(Q_1^{[k]},\dotsc,Q_N^{[k]})=Q_{1:N}^{[k]}$. As the user does not know the messages beforehand, the queries are independent of the messages, i.e.,
\begin{align}\label{3.5}
    I(W_{1:K}; Q_{1:N}^{[k]}) = 0.
\end{align}
In addition, we require that the index of the retrieved message by the user is private against any $T$ colluding databases or less, i.e., for any user-required message index $k\in\{1,\dotsc,K\}$, 
\begin{align} \label{4}
    I(k; Q_{\mathcal{T}}^{[k]},S_{\mathcal{T}}) = 0.
\end{align}
where $\mathcal{T}$ denotes any set of $T$ colluding databases among the $N$ databases and $S_{\mathcal{T}}$ is the stored content of those $T$ databases. Upon receiving the query, the $n$th database replies with a deterministic answer string $A_{n}^{[k]}$ based on its received query $Q_{n}^{[k]}$ and stored data $S_{n}$, i.e.,
\begin{align} \label{5}
    H(A_{n}^{[k]}|S_{n},Q_{n}^{[k]}) = 0, \quad n \in \{1,\ldots,N\}.
\end{align}
The user must be able to decode the required message from all answers received and the transmitted queries, i.e.,
\begin{align} \label{6}
    H(W_k|A_{1:N}^{[k]},Q_{1:N}^{[k]})=0.
\end{align}
The rate $R(B_X,T,N,K)$ of any scheme satisfying (\ref{2})-(\ref{6}) is defined as the ratio between the length of the required message and the average length of the answer strings,
\begin{align} \label{rate}
    R(B_X,T,N,K)= \frac{H(W_k)}{\mathbb{E}\big[H(A_{1:N}^{[k]})\big]}.
\end{align}

\section{Main Results} \label{mainresults}

In this section, we present achievable rates and fundamental performance limits for A-XSTPIR. The achievable scheme is based on a grouping mechanism that collects subsets of databases to perform a certain task. Throughout the rest of the paper, $M_i$ defines the number of databases in the $i$th group,\footnote{$M_i$s are problem-specific, as they are obtained by solving an optimization problem based on the given $B_X$ as discussed in Section~\ref{achievable scheme}.} $g$ defines the number of groups, and $X$ represents the maximum number of databases that are connected together based on the given set of communication links specified by $B_X$. The proposed scheme is only applicable if the following condition is satisfied by the given parameters $N$, $K$, $T$ and $B_X$,
\begin{align}\label{feasibility}
     {N \choose X} - \Omega_X(B_X) \neq 0,
\end{align}
where $\Omega_i(\cdot)$ is defined as the number of columns in a matrix with weight $i$, i.e., if $A$ is an $n \times m$ matrix $\Omega_i(A)= \sum_{j=1}^m \mathbbm{1}\Big\{\sum_{k=1}^n A(k,j) = i\Big\}$.

\begin{theorem} \label{thm3}
Consider a setting with any $T$ databases colluding to obtain information on the user-required message index and a communication matrix $B_X$ denoting all possible communications between databases that share their stored data to learn about the $K$ messages, satisfying \eqref{feasibility}. When $X+T\leq N$, the following rate is achievable with group sizes $M_i$, $i\in\{1,\dotsc,g\}$ for any $T<g$,
\begin{align}\label{ach}
        R_{A-XSTPIR}= \frac{g}{\sum_{i=1}^g M_i} C_{TPIR}(T,g,K), 
    \end{align}
where $C_{TPIR}(T,g,K)$ denotes the capacity of PIR with $g$ databases storing $K$ messages, with up to any $T$ of them allowed to collude, given by,
\begin{align}
    C_{TPIR}(T,g,K)=\left(1+\frac{T}{g}+\dotsc+\left(\frac{T}{g}\right)^{K-1}\right)^{-1}
\end{align}
\end{theorem}

\begin{remark}\label{range remark}
    The achievable rate of the proposed scheme exceeds classical XSTPIR rate when $X$ is in the range given by, $\left\lceil N\left(1-\frac{1}{\sum_{i=1}^gM_i}(g-T)\right)-T\right\rceil \leq X \leq N-g$.
\end{remark}

\begin{theorem}\label{thm4}
    When $X+T \leq N$, the maximum achievable rate of A-XSTPIR for a given $B_X$ is upper bounded by, 
    \begin{align}\label{ub}
        R_{A-XSTPIR} \leq \frac{\lambda}{M + \sum_{i=1}^M \eta(\mathcal{X}_i)},
    \end{align}
    where $\lambda=\max_{n} \sum_{m=1}^M (1-B_X(n,m)) $, $M$ is the number of columns in $B_X$, $\mathcal{X}_i$ is the set of databases in the $i$th communication link, and $\eta(\cdot)$ is the set function defined as,
    \begin{align}\label{p(x)}
        \eta(\mathcal{X}_i)\!=\! \frac{T}{N\!-\!|\mathcal{X}_i|} \!+\!\left(\!\frac{T}{N\!-\!|\mathcal{X}_i|}\!\right)^2\!+\! \dotsc \!+\! \left(\!\frac{T}{N\!-\!|\mathcal{X}_i|}\!\right)^{K\!-\!1}
    \end{align}
    where $|\cdot|$ represent the cardinality of a set.
\end{theorem}

\begin{remark}\label{equality remark}
    The rate of the proposed scheme when $|\mathcal{X}_i|=N-g$ for all $i \in \{1,\dotsc,M\}$ and $\frac{\lambda}{M} = \frac{g}{\sum_{i=1}^g M_i}$ achieves the upper bound in Theorem~\ref{thm4}.
\end{remark}

\section{Achievable Scheme} \label{achievable scheme}

In this section, we provide a database grouping mechanism followed by a group-based PIR scheme to perform A-XSTPIR. The main idea of this scheme is to combine multiple databases into groups that act as super-databases in the classical PIR setting defined in \cite{c_pir}. The goal of the grouping mechanism is to maximize the number of groups, i.e., number of super-databases, since the rate increases with the number of databases in \cite{c_pir}. At the same time, the individual databases must be carefully grouped such that the communicating/colluding databases obtain no information about the stored messages or the user-required message index.

We define a similarity matrix $S$ of size $\left\lceil \frac{N}{2}\right\rceil \times N$ to denote the grouping structure of the $N$ databases. The notation is such that $S(i,k)=S(i,j)=1$ with zero at all other entries in the $i$th row of $S$ indicates that the $k$th and $j$th databases are in the same group ($i$th group). The optimization problem that finds the optimum grouping structure under given communications between databases while satisfying all privacy and security constraints is given by,
\begin{align}\label{main_opt}
    \max_{S} & \quad \sum_{i} \mathbbm{1}\{S_i \neq \mathbf{0}_{1\times N}\}\nonumber\\
    \text{s.t.} &\quad S B^c_{X} \geq \mathbf{1}_{\frac{N}{2} \times M} \nonumber\\
    &\quad S(i,j) \in \{0,1\}, \quad  j=1,\ldots,N,\quad i=1,\dotsc,\left\lceil \frac{N}{2}\right\rceil \nonumber\\
    &\quad \sum_{j=1}^{N} S(i,j) \neq 1, \quad i=1,2,\ldots,\left\lceil \frac{N}{2}\right\rceil, \nonumber\\
    &\quad S_i^T S_j = 0, \quad i\neq j.
\end{align}
where $S_i$ is the $i$th row in $S$, $\mathbf{0}$ is the all zeros vector, $B^c_X$ is the binary complement of $B_X$ and $\mathbf{1}$ is the all ones matrix. The goal of the optimization problem in \eqref{main_opt} is to maximize the number of groups based on the communication matrix $B_X$. The first constraint is derived from $S_iB_X\leq (\sum_{j=1}^{N} S(i,j)-1)\mathbf{1}_M^T$ for each $i$, which ensures that the communicating databases in each group are unable to decode the stored messages. The third constraint avoids forming groups with a single database as we require security against each individual database, and the last constraint prevents a given database from being assigned to more than one group.\footnote{The complete description of the scheme in Section~\ref{achievable scheme} clarifies the constraints in \eqref{main_opt} better.} The solution to \eqref{main_opt} determines the optimal grouping structure among the databases that maximizes the rate while satisfying all privacy and security constraints, based on the PIR scheme described in Section~\ref{scheme}.

The optimization problem in \eqref{main_opt} is solvable only when \eqref{feasibility} is satisfied, which states the admissible set of system parameters for the proposed scheme to work. Moreover, the number of groups $g$ for a given setting is upper bounded as, 
\begin{align}
\label{groupsupperbound}
    g \leq \sum_{i=2}^X {N \choose i} - \Omega_i(B_X),
\end{align}
which decreases the dimensionality of the optimization variable defined in (\ref{main_opt}).

Note that \eqref{main_opt} maximizes the number of groups, i.e., minimizes the number of databases in each group. The reason for choosing the least number of databases within a group is stated in the following lemma, which is explained within an example in Section~\ref{eg1} and is proven in Section \ref{proofs}.

\begin{lemma} \label{lem1}
When $N$ is fixed, for any group $i$ with $M_i=dn_i$, for given constants $d$ and $n_i$, the rate of retrieval from this group is the same even if $M_i$ is increased to $M_i=dn_i+\ell$, $\ell \in \{1,2,\ldots,d-1\}$.
\end{lemma}

\subsection{Generalized Group-Based PIR Scheme}\label{scheme}

In this subsection, we describe the group-wise PIR scheme that specifies the queries sent to each database once the grouping structure is determined by solving \eqref{main_opt}. When $m$ databases are grouped together based on the optimization problem defined in \eqref{main_opt}, $m-1$ noise terms, which are generated uniformly at random from $\mathbb{F}_q$ are distributed among the databases. Database $i$, $i\in\{1,\dotsc,m-1\}$, is assigned the noise matrix $N_i$ of size $K \times L$. The noise matrix assigned for database $m$ is given by $\sum_{i=1}^{m-1} N_i$. What is stored in each database $i$, $i\in\{1,\dotsc,m-1\}$, is given by $S_i=W_{1:K}+N_i$, where $W_{1:K}$ is the $K\times L$ matrix containing the $K$ messages. Database $m$ stores the noise matrix $S_m=\sum_{i=1}^{m-1} N_i$. The user downloads noise-added information from all $m$ databases, and is able to decode the noise-free message symbols by cancelling out the $m-1$ noise terms. However, the databases within the group are unable to obtain the noise-free message symbols as there can only be up to $m-1$ communicating databases in the group based on the first constraint in \eqref{main_opt}.

To describe the general group-based PIR scheme, assume that the $N$ databases are grouped into $g$ groups where the $i$th group has $M_i$ databases. Assume that $T=1$, and all arguments provided for this case can be extended for any $T$-colluding setting using MDS codes \cite{colluding}. The PIR scheme we use in this work is identical to the scheme in \cite{c_pir} with the individual databases in \cite{c_pir} replaced by the groups in the proposed scheme. A subpacket, i.e., a block of bits, of each message consists of $g^K$ bits given by $U_i = (u_i(1),u_i(2),\ldots,u_i(g^K))$, $i \in \{1,2,\ldots, K\}$. The user obtains the $g^K$ bits of the required message using the same mechanism described in \cite{c_pir}. However, each individual database does not store the exact message bits as in \cite{c_pir} in the proposed scheme due to the additional security constraints. Therefore, the user must send the same queries to each of the $M_i$ databases in each group $i$ to decode the message bits, as there are $M_i-1$ noise terms involved in the storage within each group $i$. 

The structure of the query sent to each database is as follows. An $i$-\emph{sum} is defined as the sum of $i$ elements from $i$ distinct messages. If they were drawn from messages $j_1,j_2,\ldots,j_i$, we call it an $i$-\emph{sum} of type $\{j_1,j_2,\ldots,j_i\}$. The query sent to each database is the union of $K$  disjoint blocks, each containing $i$-\emph{sums} for $i \in \{1,2,\ldots,K\}$. Each block $i$ contains all possible types of $i$-\emph{sums}, and contains exactly $(g-1)^{i-1}$ instances of each type. 

To calculate the rate $R(B_X,T,N,K)$ of the proposed scheme, we consider the downloads corresponding to a single subpacket. Let $|A_i^{k}|$ denote the length of the answers received from the $i$th group of databases to retrieve $W_k$. Then,
\begin{align}
    |A_i^{k}| &=M_i \sum_{i=1}^{K} (g-1)^{i-1} {K\choose i}\\
    &=M_i \Big( g^{K-1} + \frac{1}{g-1}(g^{K-1}-1)\Big).
\end{align}

First, consider the case of no collusion, i.e., $T=1$. The rate of the proposed scheme for this case is computed as,
\begin{align}
    R(&B_X,1,N,K) \nonumber\\
    &= \frac{g^K}{\sum_{i=1}^g\big|A_i^{[k]}\big|}\\
    &= \frac{g^K}{(M_1+\ldots+M_g) \Big( g^{K-1} + \frac{1}{g-1}(g^{K-1}-1)\Big)}\\
    &= \frac{g}{\sum_{i=1}^g M_i} \left(1+\frac{1}{g}+\ldots+\frac{1}{g^{K-1}}\right)^{-1}\\
    & = \frac{g}{\sum_{i=1}^g M_i}C_{PIR}(g,K).
\end{align}

Now, consider the case $1 < T < g$. The database groups are obtained by solving \eqref{main_opt}. Each group is equivalent to a single database in \cite{colluding}, and the content stored in a given database in \cite{colluding} is replicated in all $M_i$ databases in group $i$, along with added noise as described previously in this section. By sending the same query to all $M_i$ databases in each group $i$, the user is able to decode the noise-free parameters from each group, and carry out PIR as in \cite{colluding}. The rate is computed as,
\begin{align}
    R(&B_X,T,N,K)  \nonumber \\
    & =\frac{g\sum_{i=1}^K (g-K)^{i-1}T^{K-i}{K-1 \choose i-1}}{\sum_{j=1}^{g}\sum_{i=1}^K M_j(g-K)^{i-1}T^{K-i}{K \choose i}}\\
    & = \frac{g}{\sum_{i=1}^g M_i} C_{TPIR}(T,g,K).
\end{align}

\begin{remark}
    In the case of $X+T \geq N$, if $X=N-1$ for any $T$ such that $X+T \geq N$, we can use the same grouping approach and have better rates compared to classical XSTPIR, if grouping is possible. The resulting rate is $R = \frac{1}{M_1 K} > \frac{1}{NK}$, where $M_1$ is the number of utilized databases, as there is only one group when $X=N-1$.
\end{remark}

\section{Examples} \label{examples}

In this section, we present some examples to explain the achievable scheme in detail. Without loss of generality, we assume that the users want to retrieve message $W_1$ and that the indices of the symbols in each message are permuted based on a permutation function chosen by the user uniformly at random, which ensures that no information is leaked by the indices of message symbols requested by the user. 

\subsection{Example 1}\label{eg1}

In this example, $N=4$, $K=2$, $T=1$, and assume that all single-way communications need to be secure, i.e., each database is not allowed to decode any of the messages. In addition, assume that database $1$ communicates with database $2$, denoted by $ 1 \leftrightarrow 2$. Thus, the communication matrix is $B_X=(1,1,0,0)^T$. In the proposed approach, the solution for the optimization problem defined in \eqref{main_opt} results in grouping database $1$ with database $3$ and database $2$ with database $4$. The stored contents in databases $1$ and $3$ are given by $S_1=W_{1:2}+N_1$ and $S_3=N_1$, respectively. Similarly, the stored contents in databases $2$ and $4$ are given by $S_2=W_{1:2}+N_2$ and $S_4=N_2$. The subpacketization of this scheme is $L=g^2=4$, and the group-wise retrieval scheme is shown in Table~\ref{scheme_eg1}.

\begin{table}[h]
\begin{center}
\begin{tabular}{ |c|c| }
\hline
  DB1-DB3& DB2-DB4\\
 \hline
 $a_1$  & $a_2$ \\  \hline
 $b_1$  & $b_2$ \\  \hline
 $a_3+b_2$ & $a_4+b_1$ \\  \hline
\end{tabular}
\end{center}
\vspace*{0.1cm}
\caption{Group-wise retrieval scheme for example $1$.}
\label{scheme_eg1}
\vspace{-0.2cm}
\end{table}

The exact downloads from each individual database in the first group is shown in Table~\ref{scheme_eg1_2}. Note that the user can decode the noise-free message bits $a_1,b_1,a_3+b_2$ using the noisy downloads from databases $1$ and $3$. The rate achieved by the proposed scheme for this example is $R=\frac{4}{12}=\frac{1}{3}$, which is greater than that of classical XSTPIR for this example, given by, $R_{XSTPIR}=1-\frac{X+T}{N}=\frac{1}{4}$.

\begin{table}[h]
\begin{center}
\begin{tabular}{ |c|c|  }
 \hline
 DB1& DB3 \\
 \hline
 $a_1+N_1(1,1)$ & $N_1(1,1)$ \\  \hline 
 $b_1+N_1(2,1)$ & $N_1(2,1)$ \\  \hline 
 $a_3+b_2+N_1(1,3)+N_1(2,2)$  & $N_1(1,3)+N_1(2,2)$ \\  \hline
\end{tabular}
\end{center}
\vspace{0.1cm}
\caption{Retrieval scheme for example $1$.}
\label{scheme_eg1_2}
\vspace{-0.2cm}
\end{table}

To see how Lemma~\ref{lem1} applies and why this is the optimal grouping, assume that we include database 4 with the first group. In the first group, we still need to download 4 symbols, two for $a_1$ and two for $b_1$ to be able to decode $a_1$ and $b_1$ to achieve privacy. On the other hand, group 2 cannot be used since it contains only one database which is not secure by itself, which deteriorates the overall rate.

\subsection{Example 2}

In this example, we show that we can drop databases to achieve better rates. Consider an example with $N=7$, $K=2$, $T=1$. The communication matrix $B_X$ is given by,
\begin{align}
    B_{X}=
    \begin{bmatrix}
        1&1&1&0&0&0&0&0&1\\
        1&0&0&1&1&0&0&0&1\\
        1&0&1&1&0&0&0&0&0\\
        0&0&0&1&1&1&1&0&0\\
        0&1&1&0&0&1&0&1&0\\
        0&1&0&0&1&1&0&0&1\\
        0&0&0&0&0&0&1&1&1
    \end{bmatrix}.
\end{align}
The solution to the optimization problem in \eqref{main_opt} is given by,
\begin{align}
    S=
    \begin{bmatrix}
        1&0&0&1&0&0&0\\
        0&1&0&0&1&0&0\\
        0&0&1&0&0&1&0\\
        0&0&0&0&0&0&0
    \end{bmatrix},
\end{align}
i.e., it groups databases $1$ with database $4$; database $2$ with database $5$; database $3$ with database $6$; and database $7$ is left alone and neglected. The subpacketization of each message is $L=g^2=9$. The group-wise retrieval scheme is shown in Table~\ref{scheme_eg2}. Within each group, the queries sent to each database take the same noise-added form described in Example~1.

\begin{table}[h]
\begin{center}
\begin{tabular}{ |c|c|c|c|  }
\hline
DB1-DB4 & DB2-DB5 & DB3-DB6 & DB7\\
 \hline
 $a_1$   & $a_2$    & $a_3$ & x\\  \hline
 $b_1$   & $b_2$    & $b_3$ & x\\  \hline
 $a_4+b_2$   & $a_5+b_1$    & $a_6+b_1$& x\\  \hline
 $a_7+b_3$   & $a_8+b_3$    & $a_9+b_2$ & x\\  \hline
\end{tabular}
\end{center}
\vspace{0.1cm}
\caption{Group-wise retrieval scheme for example $2$.}
\label{scheme_eg2}
\vspace{-0.2cm}
\end{table}

Since the maximum number of communication links for this example is $4$ the rate achieved by classical XSTPIR is $R_{XSTPIR}=1-\frac{5}{7}=\frac{2}{7}$, which is less than that of the proposed A-XSTPIR scheme given by $R =\frac{9}{24}=\frac{3}{8}$. 

\subsection{Example 3}
In this example, $N=6$, $K=3$, $T=2$, and let the communication matrix be given as, 
\begin{align}
    B_{X}=
    \begin{bmatrix}
    0&1&1&1&1&0&0\\
    1&0&0&0&0&1&1\\
    0&1&1&0&0&1&1\\
    1&0&0&1&1&0&0\\
    0&1&0&1&0&1&0\\
    0&0&1&0&1&0&1
    \end{bmatrix},
\end{align}
Based on the optimization problem defined in (\ref{main_opt}), the optimal solution is
\begin{align}
    S =
    \begin{bmatrix}
    1&1&0&0&0&0\\
    0&0&1&1&0&0\\
    0&0&0&0&1&1\\
    \end{bmatrix},
\end{align}
i.e., the databases are grouped as, database $1$ with database $2$; database $3$ with database $4$; and database $5$ with database $6$. The subpacketization of each message is $L=g^K=3^3=27$. Define $S_1, S_2,S_3 \in \mathbb{F}^{27\times 27}$ as three random full-rank matrices chosen uniformly at random and $\text{MDS}_{Y\times Z}$ as the generator matrix of any $(Y,Z)$ MDS code, all known to both the user and the databases. As the user wants to retrieve $W_1$, the following linear combinations of parameters are computed.
\begin{align}
   a_{[1:27]} &= S_1 W_1\\
   b_{[1:18]} &= \text{MDS}_{18 \times 12}S_2([1:12],:) W_2\\
   c_{[1:18]} &= \text{MDS}_{18 \times 12}S_3([1:12],:) W_3\\
   b_{[19:27]}&= \text{MDS}_{9 \times 6}S_2([13:18],:) W_2\\
   c_{[19:27]} &= \text{MDS}_{9 \times 6}S_3([13:18],:) W_3
\end{align}
The group-wise retrieval scheme is shown in Table \ref{scheme_eg3}. The scheme guarantees privacy and decodability constraints based on the proofs presented in \cite{colluding}. As shown in Section~\ref{eg1}, each of the $a,b,c$ linear combinations stored in databases are added with random noise, and the individual databases in each pair grouped in this example follows the same pattern as in Table~\ref{scheme_eg1_2}. The rate of the proposed scheme for this example is $R=\frac{27}{19\times6}=\frac{9}{38}$.

\begin{table}[h]
\begin{center}
\begin{tabular}{ |c|c|c|  }
\hline
 DB1-DB2 & DB3-D4 & DB5-DB6 \\
 \hline
 $a_1,a_2,a_3,a_4$   & $a_5,a_6,a_7,a_8$    & $a_9,a_{10},a_{11},a_{12}$ \\ \hline
 $b_1,b_2,b_3,b_4$   & $b_5,b_6,b_7,b_8$    & $b_9,a_{10},b_{11},b_{12}$ \\ \hline
 $c_1,c_2,c_3,c_4$   & $c_5,c_6,c_7,c_8$    & $c_9,c_{10},c_{11},c_{12}$\\ \hline
 $a_{13}+b_{13}$   & $a_{15}+b_{15}$    & $a_{21}+b_{17}$\\ \hline
 $a_{14}+b_{14}$   & $a_{16}+b_{16}$    & $a_{22}+b_{18}$ \\ \hline
 $a_{17}+c_{13}$   & $a_{19}+c_{15}$    & $a_{23}+c_{17}$\\ \hline
 $a_{18}+b_{14}$   & $a_{20}+c_{16}$    & $a_{24}+c_{18}$\\ \hline
 $b_{19}+c_{19}$   & $b_{21}+c_{21}$    & $b_{23}+c_{23}$\\ \hline
 $b_{20}+c_{20}$   & $b_{22}+c_{22}$    & $b_{24}+c_{24}$\\ \hline
 $a_{25}+b_{25}+c_{25}$   & $a_{26}+b_{26}+c_{26}$    & $a_{27}+b_{27}+c_{27}$\\ \hline
\end{tabular}
\end{center}
\vspace{0.1cm}
\caption{Group-wise retrieval scheme for example $3$.}
\label{scheme_eg3}
\vspace{-0.3cm}
\end{table}

\subsection{Example 4}

In this example, we show the adverse impact of solving problems with asymmetric security requirements using symmetric assumptions. Consider $N=30$ and $T=1$ with arbitrary number of messages $K$. Assume that the system has some highly communicative databases, resulting in $X=19$, i.e., there are $19$ highly communicative databases. The classical (under \emph{symmetric} assumptions) XSTPIR rate (asymptotic capacity) is $R_{XSTPIR}=1-\frac{19+1}{30}=\frac{1}{3}$. On the other hand, under \emph{asymmetric} assumptions, the solution of the optimization problem given in \eqref{main_opt} for this example results in the following grouping: $M_1=M_2=\ldots=M_{10}=2$. The asymptotic rate achieved by the proposed A-XSTPIR scheme for this system is $\lim_{K \rightarrow \infty} R = \frac{10}{20}C_{TPIR}(1,10,K)= \frac{9}{20}$. The A-XSTPIR scheme achieves a significantly higher (\%35 higher) rate even after dropping 10 databases. An interesting feature of the proposed scheme is that it uses a systematic approach to select which and how many databases to drop, to achieve the highest rate. For example, solving \eqref{main_opt} for this example shows that only $10$ out of the $19$ communication-intense databases must be dropped to achieve the highest rate.

\section{Conclusions} \label{conclusions}
In this paper, we studied the problem of asymmetric XSTPIR, which specifies all communications among the databases in a given PIR system. We proposed a database grouping mechanism followed by a group-based PIR scheme that achieves higher rates compared to classical XSTPIR in database systems with asymmetric communications. The proposed approach provides a systematic method to remove communication-intensive databases in certain cases to achieve better rates. Thus, we showed that having more databases is not always better with asymmetric database communications.

\section{Proofs}\label{proofs}
In this section, we provide proofs of Theorem~\ref{thm4}, Remark~\ref{range remark}, Remark~\ref{equality remark}, and Lemma~\ref{lem1}. 

\begin{Proof} \textbf{[Proof of Remark~\ref{range remark}]}
    As $R_{XSTPIR}=1-\frac{X+T}{N}$ is decreasing in $X$, showing that,
    \begin{align}\label{33}
        X=N\Big(1- \frac{1}{\sum_{i=1}^g M_i}(g-T)\Big)-T
    \end{align}
    satisfies $R_{XSTPIR}=\lim_{K\rightarrow\infty}R_{A-XSTPIR}$ proves the lower bound on $X$ for any arbitrary $K$. Therefore, for the value of $X$ in \eqref{33}, $R_{XSTPIR}$ is given by, 
    \begin{align}
    1\!-\!\frac{X\!+\!T}{N} \!=\! \frac{g}{\sum_{i=1}^g M_i}\left(1\!-\!\frac{T}{g}\right)\!=\!\lim_{K\rightarrow\infty} R_{A-XSTPIR} 
    \end{align}
which concludes the proof. 
\end{Proof}

\begin{Proof}\textbf{[Proof of Theorem~\ref{thm4}]}
    We start by using \cite[Lem.~3]{csa} and \cite[(89)]{csa}, which are valid for any set of $X$ communicating databases in a general XSTPIR setting. Considering any arbitrary communication link $\mathcal{X}_i$ from the given $B_X$, the combination of \cite[Lem.~3]{csa} and \cite[(89)]{csa} can be written as,
    \begin{align}
        L \leq \sum_{n \in \bar{\mathcal{X}_i}}D_n - L \eta(\mathcal{X}_i), 
    \end{align}
    where $\bar{\mathcal{X}_i}$ is the complement of the set $\mathcal{X}_i$, $D_n$ is the expected number of downloads from database $n$, and $\eta(\mathcal{X}_i)$ is as defined in \eqref{p(x)}. Summing across all $M$ links we get,
    \begin{align}
        \!\!\! ML \leq \sum_{i=1}^M \sum_{n \in \bar{\mathcal{X}_i}}D_n - L\sum_{i=1}^M \eta(\mathcal{X}_i)\leq \lambda D -  L \sum_{i=1}^M \eta(\mathcal{X}_i)\label{lambdawhy}
    \end{align}
    where $\lambda=\max_{n} \sum_{m=1}^M (1-B_X(n,m))$ is the maximum row sum of $B_X^c$ and $D=\sum_{n=1}^N D_n$ is the total download. Note that \eqref{lambdawhy} is a typical over counting argument, and the proof is completed by substituting $R=\frac{L}{D}$ in \eqref{lambdawhy}.
\end{Proof}

\begin{Proof}\textbf{[Proof of Remark~\ref{equality remark}]}
    Substituting $|\mathcal{X}_i|=N-g$ and $\frac{\lambda}{M}=\frac{g}{N}$ in \eqref{ub} of Theorem~\ref{thm4} gives the achievable rate \eqref{ach} as the upper bound on $R_{A-XSTPIR}$ for this case.
\end{Proof}

\begin{Proof}\textbf{[Proof of Lemma~\ref{lem1}]} \label{lemma3proof}
We prove this lemma using induction and for the case of $d=2$, i.e., we need only two copies of the symbol to be able to decode it. For arbitrary $d$, it is held mutates mutandis, thus the proof generalizes, and the case of $2$ is sufficient to grasp the important concepts. Assume that it is required to retrieve $a_\ell$ and $a_j$, $\ell \neq j$, and that we have side information $b_k$ from previous steps. Along the proof, the noise terms are dropped for simplicity.

Step~1: Assume $n_i$ = 1. When we have $M_i=2n_i+1=3$. The minimum amount of downloads required to retrieve $a_\ell$ and $a_j$ privately given $b_k$ as side information is $6$ symbols. The only possible retrieval structure using $3$ databases is to retrieve $a_\ell$, $b_m$, and $a_j+b_k$ from one of the databases, $a_\ell$ and $b_m$ from the second database, and finally $a_j+b_k$ from the third database. It is clear that we can reduce this retrieval scheme to achieve the same rate using only $2$ databases, i.e., $M_i=2$, by retrieving $a_\ell$, $b_m$, and $a_j+b_k$ from the first and second databases. 
Step~2: Assume that this holds for $n_i = n$.
Step~3: $n_i = n+1$. In this case, we have $M_i = 2 n+3 $, it is clear that we can further subdivide $M_i$ into $n$ groups of $2$ and one group of $3$ and still achieve the same rate. For the group of $3$, we have the same argument as in Step~1.
\end{Proof}

\bibliographystyle{unsrt}
\bibliography{references.bib}

\begin{thebibliography}{10}

\bibitem{chor}
B.~Chor, E.~Kushilevitz, O.~Goldreich, and M.~Sudan.
\newblock Private information retrieval.
\newblock {\em Jour. of the ACM}, 45(6):965--981, November 1998.

\bibitem{c_pir}
H.~Sun and S.~A. Jafar.
\newblock The capacity of private information retrieval.
\newblock {\em IEEE Trans. Info. Theory}, 63(7):4075--4088, July 2017.

\bibitem{c_spir}
H.~Sun and S.~A. Jafar.
\newblock The capacity of symmetric private information retrieval.
\newblock {\em IEEE Trans. Info. Theory}, 65(1):322--329, June 2018.

\bibitem{wang_spir}
Z.~Wang and S.~Ulukus.
\newblock Symmetric private information retrieval at the private information
  retrieval rate.
\newblock {\em IEEE Jour. on Selected Areas in Info. Theory}, 3(2):350--361,
  June 2022.

\bibitem{tspir_mdscoded}
Q.~Wang and M.~Skoglund.
\newblock Symmetric private information retrieval from {MDS} coded distributed
  storage with non-colluding and colluding servers.
\newblock {\em IEEE Trans. Info. Theory}, 65(8):5160--5175, March 2019.

\bibitem{banawan_pir_mdscoded}
K.~Banawan and S.~Ulukus.
\newblock The capacity of private information retrieval from coded databases.
\newblock {\em IEEE Trans. Info. Theory}, 64(3):1945--1956, January 2018.

\bibitem{codedstorage_adversary_tpir}
L.~Holzbaurand, R.~Freij-Hollanti, and C.~Hollanti.
\newblock On the capacity of private information retrieval from coded,
  colluding, and adversarial servers.
\newblock In {\em IEEE ITW}, August 2019.

\bibitem{Salim_CodedPIR}
R.~Tajeddine, O.~Gnilke, and S.~El Rouayheb.
\newblock Private information retrieval from {MDS} coded data in distributed
  storage systems.
\newblock {\em IEEE Trans. Info. Theory}, 64(11):7081--7093, March 2018.

\bibitem{Kumar_PIRarbCoded}
S.~Kumar, H.-Y. Lin, E.~Rosnes, and A.~G. i~Amat.
\newblock Achieving maximum distance separable private information retrieval
  capacity with linear codes.
\newblock {\em IEEE Trans. Info. Theory}, 65(7):4243--4273, July 2019.

\bibitem{banawan_multimessage_pir}
K.~Banawan and S.~Ulukus.
\newblock Multi-message private information retrieval: Capacity results and
  near-optimal schemes.
\newblock {\em IEEE Trans. Info. Theory}, 64(10):6842--6862, April 2018.

\bibitem{multimessage_pir_sideinfo}
M.~J. Siavoshani, S.~P. Shariatpanahi, and M.~Ali Maddah-Ali.
\newblock Private information retrieval for a multi-message scenario with
  private side information.
\newblock {\em IEEE Trans. on Commun.}, 69(5):3235--3244, January 2021.

\bibitem{colluding}
H.~Sun and S.~A. Jafar.
\newblock The capacity of robust private information retrieval with colluding
  databases.
\newblock {\em IEEE Trans. Info. Theory}, 64(4):2361--2370, April 2018.

\bibitem{byzantine_tpir}
K.~Banawan and S.~Ulukus.
\newblock The capacity of private information retrieval from {B}yzantine and
  colluding databases.
\newblock {\em IEEE Trans. Info. Theory}, 65(2):1206--1219, September 2018.

\bibitem{uncoded_constrainedstorage_pir}
M.~A. Attia, D.~Kumar, and R.~Tandon.
\newblock The capacity of private information retrieval from uncoded storage
  constrained databases.
\newblock {\em IEEE Trans. Info. Theory}, 66(11):6617--6634, September 2020.

\bibitem{batuhan_hetero}
K.~Banawan, B.~Arasli, Y.-P. Wei, and S.~Ulukus.
\newblock The capacity of private information retrieval from heterogeneous
  uncoded caching databases.
\newblock {\em IEEE Trans. Info. Theory}, 66(6):3407--3416, June 2020.

\bibitem{utah_hetero}
N.~Woolsey, R.~Chen, and M.~Ji.
\newblock Uncoded placement with linear sub-messages for private information
  retrieval from storage constrained databases.
\newblock {\em IEEE Trans. Commun.}, 68(10):6039--6053, October 2020.

\bibitem{wei_banawan_cache_pir}
Y.-P. Wei, K.~Banawan, and S.~Ulukus.
\newblock Fundamental limits of cache-aided private information retrieval with
  unknown and uncoded prefetching.
\newblock {\em IEEE Trans. Info. Theory}, 65(5):3215--3232, November 2018.

\bibitem{chaoTian}
C.~Tian, H.~Sun, and J.~Chen.
\newblock Capacity-achieving private information retrieval codes with optimal
  message size and upload cost.
\newblock {\em IEEE Trans. on Info. Theory}, 65(11):7613--7627, November 2019.

\bibitem{semantic_pir}
S.~Vithana, K.~Banawan, and S.~Ulukus.
\newblock Semantic private information retrieval.
\newblock {\em IEEE Trans. Info. Theory}, 68(4):2635--2652, December 2021.

\bibitem{salim_singleserver_pir}
A.~Heidarzadeh, S.~Kadhe, S.~El Rouayheb, and A.~Sprintson.
\newblock Single-server multi-message individually-private information
  retrieval with side information.
\newblock In {\em IEEE ISIT}, July 2019.

\bibitem{first_xsecure}
H.~Yang, W.~Shin, and J.~Lee.
\newblock Private information retrieval for secure distributed storage systems.
\newblock {\em IEEE Trans. Info. Foren. Security}, 13(12):2953--2964, May 2018.

\bibitem{csa}
Z.~Jia, H.~Sun, and S.~A. Jafar.
\newblock Cross subspace alignment and the asymptotic capacity of {$X$}-secure
  {$T$}-private information retrieval.
\newblock {\em IEEE Trans. Info. Theory}, 65(9):5783--5798, May 2019.

\bibitem{nan_eaves}
J.~Cheng, N.~Liu, W.~Kang, and Y.~Li.
\newblock The capacity of symmetric private information retrieval under
  arbitrary collusion and eavesdropping patterns.
\newblock {\em IEEE Trans. Info. Foren. Security}, 17:3037--3050, August 2022.

\end{thebibliography}
\end{document}